\pdfoutput=1
\documentclass[10pt,a4paper,onecolumn]{article}
\usepackage{marginnote}
\usepackage{graphicx}
\usepackage{xcolor}
\usepackage{authblk,etoolbox}
\usepackage{titlesec}
\usepackage{calc}
\usepackage{tikz}
\usepackage{hyperref}
\hypersetup{colorlinks,breaklinks=true,
            citecolor=[rgb]{0.1, 0.1, 0.6},
            urlcolor=[rgb]{0.0, 0.5, 1.0},
            linkcolor=[rgb]{0.0, 0.5, 1.0}}
\usepackage{caption}
\usepackage{tcolorbox}
\usepackage{amssymb,amsmath}
\usepackage{ifxetex,ifluatex}
\usepackage{seqsplit}
\usepackage{xstring}
\usepackage{float}
\let\origfigure\figure
\let\endorigfigure\endfigure
\renewenvironment{figure}[1][2] {
    \expandafter\origfigure\expandafter[H]
} {
    \endorigfigure
}

\usepackage{natbib}
\setcitestyle{authoryear,open={(},close={)}}

\usepackage{pgfplots}
\usepackage{pgf}
\usepackage{tikzscale}
\usepackage{tikz}
\pgfplotsset{compat=newest}
\usetikzlibrary{arrows,calc,positioning,shapes.geometric,math}

\usepackage{amsmath, amssymb, amsthm}
\DeclareMathOperator{\Conv}{Conv1d}
\DeclareMathOperator{\Convtwo}{Conv2d}
\DeclareMathOperator{\Convtr}{ConvTr1d}
\DeclareMathOperator{\ReLU}{Relu}
\DeclareMathOperator{\GELU}{GELU}
\DeclareMathOperator{\GLU}{GLU}
\DeclareMathOperator{\SDR}{SDR}
\DeclareMathOperator{\LSTM}{BiLSTM}
\DeclareMathOperator{\LN}{LN}
\def\pscale{1}
\def\trapangle{75}
\newdimen\blockh
\let\textttOrig=\texttt
\def\texttt#1{\expandafter\textttOrig{\seqsplit{#1}}}
\renewcommand{\seqinsert}{\ifmmode
  \allowbreak
  \else\penalty6000\hspace{0pt plus 0.02em}\fi}

\makeatletter
\let\href@Orig=\href
\def\href@Urllike#1#2{\href@Orig{#1}{\begingroup
    \def\Url@String{#2}\Url@FormatString
    \endgroup}}
\def\href@Notdoi#1#2{\def\tempa{#1}\def\tempb{#2}%
  \ifx\tempa\tempb\relax\href@Urllike{#1}{#2}\else
  \href@Orig{#1}{#2}\fi}
\def\href#1#2{%
  \IfBeginWith{#1}{https://doi.org}%
  {\href@Urllike{#1}{#2}}{\href@Notdoi{#1}{#2}}}
\makeatother

\usepackage[top=3.5cm, bottom=3cm, right=1.5cm, left=1.0cm,
            headheight=2.2cm, reversemp, includemp, marginparwidth=4.5cm]{geometry}

\titleformat{\section}
  {\normalfont\sffamily\Large\bfseries}
  {}{0pt}{}
\titleformat{\subsection}
  {\normalfont\sffamily\large\bfseries}
  {}{0pt}{}
\titleformat{\subsubsection}
  {\normalfont\sffamily\bfseries}
  {}{0pt}{}
\titleformat*{\paragraph}
  {\sffamily\normalsize}

\usepackage{fancyhdr}
\makeatletter
\let\ps@plain\ps@fancy
\fancyheadoffset[L]{4.5cm}
\fancyfootoffset[L]{4.5cm}

\definecolor{linky}{rgb}{0.0, 0.5, 1.0}

\newtcolorbox{repobox}
   {colback=red, colframe=red!75!black,
     boxrule=0.5pt, arc=2pt, left=6pt, right=6pt, top=3pt, bottom=3pt}

\patchcmd{\@maketitle}{center}{flushleft}{}{}
\patchcmd{\@maketitle}{center}{flushleft}{}{}
\patchcmd{\@maketitle}{\LARGE}{\LARGE\sffamily}{}{}
\def\maketitle{{%
  
  \AB@maketitle}}
\makeatletter
\renewcommand\AB@affilsepx{ \protect\Affilfont}
\renewcommand\AB@affilnote[1]{{\bfseries #1}\hspace{3pt}}
\renewcommand{\affil}[2][]%
   {\newaffiltrue\let\AB@blk@and\AB@pand
      \if\relax#1\relax\def\AB@note{\AB@thenote}\else\def\AB@note{#1}%
        \setcounter{Maxaffil}{0}\fi
        \begingroup
        \let\href=\href@Orig
        \let\texttt=\textttOrig
        \let\protect\@unexpandable@protect
        \def\thanks{\protect\thanks}\def\footnote{\protect\footnote}%
        \@temptokena=\expandafter{\AB@authors}%
        {\def\\{\protect\\\protect\Affilfont}\xdef\AB@temp{#2}}%
         \xdef\AB@authors{\the\@temptokena\AB@las\AB@au@str
         \protect\\[\affilsep]\protect\Affilfont\AB@temp}%
         \gdef\AB@las{}\gdef\AB@au@str{}%
        {\def\\{, \ignorespaces}\xdef\AB@temp{#2}}%
        \@temptokena=\expandafter{\AB@affillist}%
        \xdef\AB@affillist{\the\@temptokena \AB@affilsep
          \AB@affilnote{\AB@note}\protect\Affilfont\AB@temp}%
      \endgroup
       \let\AB@affilsep\AB@affilsepx
}
\makeatother

\renewcommand\Affilfont{\sffamily\small\mdseries}
\ifnum 0\ifxetex 1\fi\ifluatex 1\fi=0 
  \usepackage[T1]{fontenc}
  \usepackage[utf8]{inputenc}

\else 
  \ifxetex
    \usepackage{mathspec}
    \usepackage{fontspec}

  \else
    \usepackage{fontspec}
  \fi
  \defaultfontfeatures{Ligatures=TeX,Scale=MatchLowercase}

\fi
\IfFileExists{upquote.sty}{\usepackage{upquote}}{}
\IfFileExists{microtype.sty}{%
\usepackage{microtype}
\UseMicrotypeSet[protrusion]{basicmath} 
}{}

\usepackage{hyperref}
\hypersetup{unicode=true,
            pdftitle={Hybrid Spectrogram and Waveform Source Separation},
            pdfborder={0 0 0},
            breaklinks=true}
\usepackage{longtable,booktabs}

\let\addcontentslineOrig=\addcontentsline
\def\addcontentsline#1#2#3{\bgroup
  \let\texttt=\textttOrig\addcontentslineOrig{#1}{#2}{#3}\egroup}
\let\markbothOrig\markboth
\def\markboth#1#2{\bgroup
  \let\texttt=\textttOrig\markbothOrig{#1}{#2}\egroup}
\let\markrightOrig\markright
\def\markright#1{\bgroup
  \let\texttt=\textttOrig\markrightOrig{#1}\egroup}

\usepackage{graphicx,grffile}
\makeatletter
\def\maxwidth{\ifdim\Gin@nat@width>\linewidth\linewidth\else\Gin@nat@width\fi}
\def\maxheight{\ifdim\Gin@nat@height>\textheight\textheight\else\Gin@nat@height\fi}
\makeatother
\setkeys{Gin}{width=\maxwidth,height=\maxheight,keepaspectratio}
\IfFileExists{parskip.sty}{%
\usepackage{parskip}
}{
\setlength{\parindent}{0pt}
\setlength{\parskip}{6pt plus 2pt minus 1pt}
}
\ifx\paragraph\undefined\else
\let\oldparagraph\paragraph
\renewcommand{\paragraph}[1]{\oldparagraph{#1}\mbox{}}
\fi
\ifx\subparagraph\undefined\else
\let\oldsubparagraph\subparagraph
\renewcommand{\subparagraph}[1]{\oldsubparagraph{#1}\mbox{}}
\fi

\title{Hybrid Spectrogram and Waveform Source Separation}

        \author[1]{Alexandre Défossez}

      \affil[1]{Facebook AI Research}
  \date{\vspace{-7ex}}

\begin{document}
\maketitle

\marginpar{

  \begin{flushleft}
  \sffamily\small

  \vspace{2mm}

  \par\noindent\hrulefill\par

  \vspace{2mm}

  \vspace{2mm}
  {\bfseries License}\\
  Authors of papers retain copyright and release the work under a Creative Commons Attribution 4.0 International License (\href{http://creativecommons.org/licenses/by/4.0/}{\color{linky}{CC BY 4.0}}).

  \vspace{4mm}
  {\bfseries In partnership with}\\
  \vspace{2mm}
  \includegraphics[width=4cm]{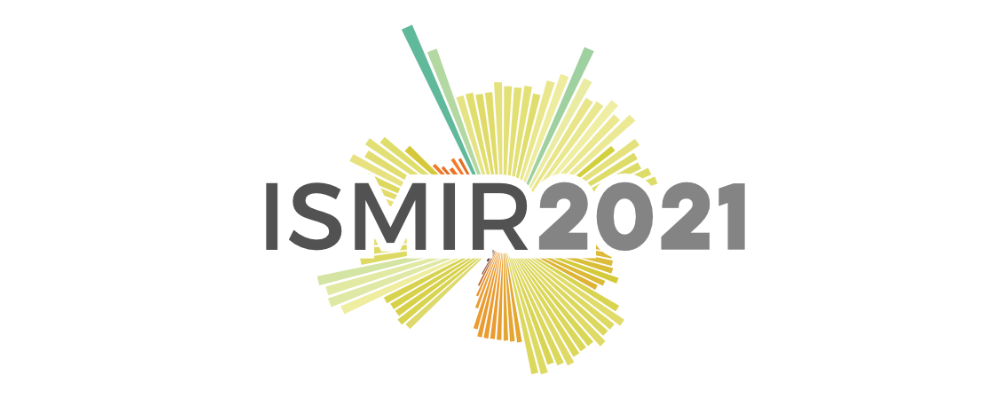}

  \end{flushleft}
}

\hypertarget{abstract}{%
\section{Abstract}\label{abstract}}

Source separation models either work on the spectrogram or waveform
domain. In this work, we show how to perform end-to-end hybrid source
separation, letting the model decide which domain is best suited for
each source, and even combining both. The proposed hybrid version of the
Demucs architecture \citep{demucs} won the Music Demixing Challenge 2021
organized by Sony. This architecture also comes with additional
improvements, such as compressed residual branches, local attention or
singular value regularization. Overall, a 1.4 dB improvement of the
Signal-To-Distortion (SDR) was observed across all sources as measured
on the MusDB HQ dataset \citep{musdbhq}, an improvement confirmed by
human subjective evaluation, with an overall quality rated at 2.83 out
of 5 (2.36 for the non hybrid Demucs), and absence of contamination at
3.04 (against 2.37 for the non hybrid Demucs and 2.44 for the second
ranking model submitted at the competition).

\hypertarget{introduction}{%
\section{Introduction}\label{introduction}}

Work on music source separation has recently focused on the task of
separating 4 well defined instruments in a supervised manner: drums,
bass, vocals and other accompaniments. Recent evaluation campaigns
\citep{sisec} have focused on this setting, relying on the standard
MusDB benchmark \citep{musdb}. In 2021, Sony organized the Music
Demixing Challenge (MDX) \citep{mdx}, an online competition where
separation models are evaluated on a completely new and hidden test set
composed of 27 tracks.

The challenge featured a number of baselines to start from, which could
be divided into two categories: spectrogram or waveform based methods.
The former consists in models that are fed with the input spectrogram,
either represented by its amplitude, such as Open-Unmix \citep{umx} and
its variant CrossNet Open-Unmix \citep{xumx}, or as the concatenation of
its real and imaginary part, a.k.a Complex-As-Channels (CAC)
\citep{cac}, such as LaSAFT \citep{lasaft}. Similarly, the output can be
either a mask on the input spectrogram, complex modulation of the input
spectrogram \citep{kong2021decoupling}, or the CAC representation.

On the other hand, waveform based models such as Demucs \citep{demucs}
are directly fed with the raw waveform, and output the raw waveform for
each of the source. Most of those methods will perform some kind of
learnt time-frequency analysis in its first layers through convolutions,
such as Demucs and Conv-TasNet \citep{convtasnet}, although some will
not rely at all on convolutional layers, like Dual-Path RNN
\citep{dual_path}.

Theoretically, there should be no difference between spectrogram and
waveform models, in particular when considering CaC (complex as
channels), which is only a linear change of base for the input and
output space. However, this would only hold true in the limit of having
an infinite amount of training data. With a constrained dataset, such as
the 100 songs of MusDB, inductive bias can play an important role. In
particular, spectrogram methods varies by more than their input and
output space. For instance, with a notion of frequency, it is possible
to apply convolutions along frequencies, while waveform methods must use
layers that are fully connected with respect to their channels. The
final test loss being far from zero, there will also be artifacts in the
separated audio. Different representations will lead to different
artifacts, some being more noticeable for the drums and bass (phase
inconsitency for spectrogram methods will make the attack sounds
hollow), while others are more noticeable for the vocals (vocals
separated by Demucs suffer from crunchy static noise)

In this work, we extend the Demucs architecture to perform hybrid
waveform/spectrogram domain source separation. The original U-Net
architecture \citep{unet} is extended to provide two parallel branches:
one in the time (temporal) and one in the frequency (spectral) domain.
We bring other improvements to the architecture, namely compressed
residual branches comprising dilated convolutions \citep{dilated}, LSTM
\citep{lstm} and attention \citep{attention} with a focus on local
content. We measure the impact of those changes on the MusDB benchmark
and on the MDX hidden test set, as well as subjective evaluations.
Hybrid Demucs ranked 1st at the MDX competition when trained only on
MusDB, with 7.32 dB of SDR, and 2nd with extra training data allowed.

\hypertarget{related-work}{%
\section{Related work}\label{related-work}}

There exist a number of spectrogram based music source separation
architectures. Open-Unmix \citep{umx} is based on fully connected layers
and a bi-LSTM. It uses multi-channel Wiener filtering \citep{wiener} to
reduce artifacts. While the original Open-Unmix is trained independently
on each source, a multi-target version exists \citep{xumx}, through a
shared averaged representation layer. D3Net \citep{d3net} is another
architecture, based on dilated convolutions connected with dense skip
connections. It was before the competition the best performing
spectrogram architecture, with an average SDR of 6.0 dB on MusDB. Unlike
previous methods which are based on masking, LaSAFT \citep{lasaft} uses
Complex-As-Channels \citep{cac} along with a U-Net \citep{unet}
architecture. It is also single-target, however its weights are shared
across targets, using a weight modulation mechanism to select a specific
source.

Waveform domain source separation was first explored by \citet{wavenet},
as well as \citet{waveunet_singing} and \citet{waveunet} with
Wave-U-Net. However, those methods were lagging in term of quality,
almost 2 dB behind their spectrogram based competitors. Demucs
\citep{demucs} was built upon Wave-U-Net, using faster strided
convolutions rather than explicit downsampling, allowing for a much
larger number of channels, but potentially introducing aliasing
artifacts as noted by \citet{pons2021upsampling}, and extra Gated Linear Unit layers
\citep{glu} and biLSTM. For the first time, waveform domain methods
surpassed spectrogram ones when considering the overall SDR (6.3 dB on
MusDB), although its performance is still inferior on the other and
vocals sources. Conv-Tasnet \citep{convtasnet}, a model based on masking
over a learnt time-frequency representation using dilated convolutions,
was also adapted to music source separation by \citet{demucs}, but
suffered from more artifacts and lower SDR.

To the best of our knowledge, no other work has studied true end-to-end
hybrid source separation, although other teams in the MDX competition
used model blending from different domains as a simpler post-training
alternative.

\hypertarget{architecture}{%
\section{Architecture}\label{architecture}}

In this Section we present the structure of Hybrid Demucs, as well as
the other additions that were added to the original Demucs architecture.

\hypertarget{original-demucs}{%
\subsection{Original Demucs}\label{original-demucs}}

The original Demucs architecture \citep{demucs} is a U-Net \citep{unet}
encoder/decoder structure. A BiLSTM \citep{lstm} is applied between the
encoder and decoder to provide long range context. The encoder and
decoder have a symetric structure. Each encoder layer is composed of a
convolution with a kernel size of 8, stride of 4 and doubling the number
of channels (except for the first layer, which sets it to a fix value,
typically 48 or 64). It is followed by a ReLU, and a so called 1x1
convolution with Gated Linear Unit activation \citep{glu}, i.e.~a
convolution with a kernel size of 1, where the first half of the
channels modulates the second half through a sigmoid. The 1x1
convolution double the channels, and the GLU halves them, keeping them
constant overall. Symetrically, a decoder layer sums the contribution
from the U-Net skip connection and the previous layer, apply a 1x1
convolution with GLU, then a transposed convolution that halves the
number of channels (except for the outermost layer), with a kernel size
of 8 and stride of 4, and a ReLU (except for the outermost layer). There
are 6 encoder layers, and 6 decoder layers, for processing 44.1 kHz
audio. In order to limit the impact of aliasing from the outermost
layers, the input audio is upsampled by a factor of 2 before entering
the encoder, and downsampled by a factor of 2 when leaving the decoder.

\hypertarget{hybrid-demucs}{%
\subsection{Hybrid Demucs}\label{hybrid-demucs}}

\hypertarget{overall-architecture}{%
\subsubsection{Overall architecture}\label{overall-architecture}}

Hybrid Demucs extends the original architecture with multi-domain
analysis and prediction capabilities. The model is composed of a
temporal branch, a spectral branch, and shared layers. The temporal
branch takes the input waveform and process it like the standard Demucs.
It contains 5 layers, which are going to reduce the number of time steps
by a factor of \(4^5 = 1024\). Compared with the original architecture,
all ReLU activations are replaced by Gaussian Error Linear Units (GELU)
\citep{gelu}.

The spectral branch takes the spectrogram obtained from a STFT over 4096
time steps, with a hop length of 1024. Notice that the number of time
steps immediately matches that of the output of the encoder of the
temporal branch. In order to reduce the frequency dimension, we apply
the same convolutions as in the temporal branch, but along the frequency
dimension. Each layer reduces by a factor of 4 the number of
frequencies, except for the 5th layer, which reduces by a factor of 8.
After being processed by the spectral encoder, the signal has only one
``frequency'' left, and the same number of channels and sample rate as
the output of the temporal branch. The temporal and spectral
representations are then summed before going through a shared
encoder/decoder layer which further reduces by 2 the number of time
steps (using a kernel size of 4). Its output serves both as the input of
the temporal and spectral decoder. Hybrid Demucs has a dual U-Net
structure, with the temporal and spectral branches having their
respective skip connections.

\begin{figure}
\hypertarget{fig:hyb}{%
\centering
\resizebox{\textwidth}{!}{
  \input{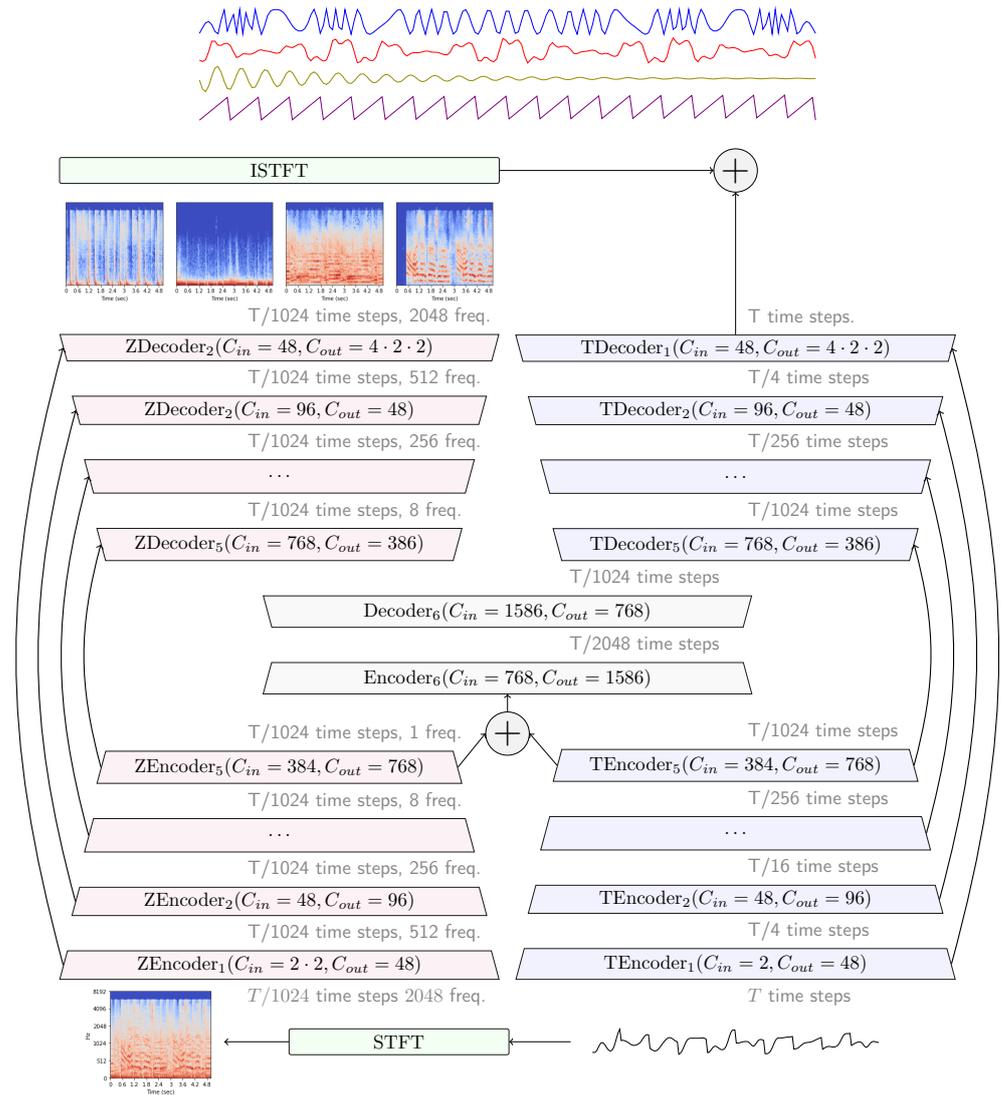}
}
\caption{Hybrid Demucs architecture. The input waveform is processed
both through a temporal encoder, and first through the STFT followed by
a spectral encoder. The two representations are summed when their
dimensions align. The decoder is built symmetrically. The output
spectrogram go through the ISTFT and is summed with the waveform
outputs, giving the final model output. The \(\mathrm{Z}\) prefix is
used for spectral layers, and \(\mathrm{T}\) prefix for the temporal
ones.}\label{fig:hyb}
}
\end{figure}

The output of the spectral branch is inversed with the ISTFT, and summed
with the temporal branch output, giving the final model prediction. Due
to this overall design, the model is free to use whichever
representation is most conveniant for different parts of the signal,
even within one source, and can freely share information between the two
representations. The hybrid architecture is represented on
\ref{fig:hyb}.

\hypertarget{padding-for-easy-alignment}{%
\subsubsection{Padding for easy
alignment}\label{padding-for-easy-alignment}}

One difficulty was to properly align the spectral and temporal
representations for any input length. For an input length \(L\), kernel
size \(K\), stride \(S\) and padding on each side \(P\), the output of a
convolution is of length \((L - K + 2 * P) / S + 1\). Following the
practice from models like MelGAN \citep{melgan} we pad by
\(P = (K - S) / 2\), giving an output of \(L / S\), so that matching the
overall stride is now sufficiant to exactly match the length of the
spectral and temporal representations. We apply this padding both for
the STFT, and convolution layers in the temporal encoders.

\hypertarget{frequency-wise-convolutions}{%
\subsubsection{Frequency-wise
convolutions}\label{frequency-wise-convolutions}}

In the spectral branch, we use frequency-wise convolutions, dividing the
number of frequency bins by 4 with every layer. For simplicity we drop
the highest bin, giving 2048 frequency bins after the STFT. The input of
the 5th layer has 8 frequency bins, which we reduce to 1 with a
convolution with a kernel size of 8 and no padding. It has been noted
that unlike the time axis, the distribution of musical signals is not
truely invariant to translation along the frequency axis. Instruments
have specific pitch range, vocals have well defined formants etc. To
account for that, \citet{zemb} suggest injecting an embedding of the
frequency before applying the convolution. We use the same approach,
with the addition that we smooth the initial embedding so that close
frequencies have similar embeddings. We inject this embedding just
before the second encoder layer. We also investigated using specific
weights for different frequency bands. This however turned out more
complex for a similar result.

\hypertarget{spectrogram-representation}{%
\subsubsection{Spectrogram
representation}\label{spectrogram-representation}}

We investigated both with representing the spectrogram as complex
numbers \citep{cac} or as amplitude spectrograms.
We also experimented with using Wiener filtering \citep{wiener}, using Open-Unmix
differentiable implementation \citep{umx}, which uses an iterative procedure.
Using more iterations at evaluation
time is usually optimal, but sadly doesn't work well with the hybrid
approach, as changing the spectrogram output, without the waveform
output being able to adapt will drastically reduce the SDR, and using a
high number of iterations at train time is prohibitively slow. In all
cases, we differentiably transform the spectrogram branch output to a
waveform, summed to the waveform branch output, and the final loss is
applied in the waveform domain.

\hypertarget{compressed-residual-branches}{%
\subsection{Compressed residual
branches}\label{compressed-residual-branches}}

The original Demucs encoder layer is composed of a convolution with
kernel size of 8 and stride of 4, followed by a ReLU, and of a
convolution with kernel size of 1 followed by a GLU. Between those two
convolutions, we introduce two compressed residual branches, composed of
dilated convolutions, and for the innermost layers, a biLSTM with
limited span and local attention. Remember that after the first
convolution of the 5th layer, the temporal and spectral branches have
the same shape. The 5th layer of each branch actually only contains this
convolution, with the compressed residual branch and 1x1 convolution
being shared.

Inside a residual branch, all convolutions are with respect to the time
dimension, and different frequency bins are processed separately. There
are two compressed residual branch per encoder layer. Both are composed
of a convolution with a kernel size of 3, stride of 1, dilation of 1 for
the first branch and 2 for the second, and 4 times less output
dimensions than the input, followed by layer normalization
\citep{layernorm} and a GELU activation.

For the 5th and 6th encoder layers, long range context is processed
through a local attention layer (see definition hereafter) as well as a
biLSTM with 2 layers, inserted with a skip connection, and with a
maximum span of 200 steps. In practice, the input is splitted into
frames of 200 time steps, with a stride of 100 steps. Each frame is
processed concurrently, and for any time step, the output from the frame
for which it is the furthest away from the edge is kept.

\begin{figure}
\hypertarget{fig:residual}{%
\centering
\resizebox{\textwidth}{!}{
  \begin{tikzpicture}[
    every node/.style={scale=\pscale},
    conv/.style={shape=trapezium,
        trapezium angle=\trapangle, draw, inner xsep=0pt, inner ysep=0pt,
        draw=black!90,fill=gray!5},
    rewrite/.style={shape=rectangle,
        draw, inner xsep=8pt, inner ysep=3pt,
        draw=black!90,fill=gray!5},
    resi/.style={shape=rectangle,
        draw, inner xsep=2pt, inner ysep=2pt, minimum width=8cm,
        draw=black!90,fill=green!5, anchor=south},
    inout/.style={rounded corners=1pt,rectangle,draw=black!90,
        fill=violet!5,minimum width=0.6cm, minimum height=0.6cm},
    skip/.style={line width=0.2mm, ->},
]
    \newcommand\curp{\the\tikz@lastxsaved,\the\tikz@lastysaved}
    \newdimen\yshift
    \newdimen\base
    \newdimen\dec
    \tikzmath{
      \yshift=0.8em;
      \base=15cm;
      \dec = 2 * \blockh / tan(\trapangle);
    }
    \def\deltax{3cm}

    \node (base) at (0, 0) {};
    \node (conv) [conv, minimum width=\base, anchor=south] at
      (base.north) {$\GELU(\Conv(C_{in}, C_{out}, K=8, S=4))$};

    \node (resi1) [resi] at ($(conv.north) + (\deltax, \yshift)$)
    {$\GELU(\LN(\Conv(C_{out}, C_{out} / 4, K=3, D=1)))$};

    \node (lstm) [resi, yshift=\yshift, dashed] at (resi1.north)
    {$\LSTM(\mathrm{layers}=2, \mathrm{span}=200)$};

    \node (attn) [resi, yshift=\yshift, dashed] at (lstm.north)
    {$\mathrm{LocalAttention}(\mathrm{heads}=4)$};

    \node [anchor=north west] at (attn.north east) {$\Bigg\}$ if $i \in \{5, 6\}$};

    \node (resi2_pre) [resi, yshift=\yshift] at (attn.north)
    {$\GLU(\LN(\Conv(C_{out} / 4, 2 \cdot C_{out}, K=1)))$};
    \node (resi2) [resi, yshift=\yshift] at (resi2_pre.north)
    {$\mathrm{LayerScale}(\mathrm{init}=1\mathrm{e}{-}3)$};

    \draw [->] (conv.north -| resi1.south) -- (resi1.south);
    \draw [->] (resi1.north)  to node (smid){} (lstm.south);
    \draw[->] (lstm.north) to node (nmid){} (attn.south);
    \draw[-] (smid.center) -- (smid -| lstm.west) -- ++ (-0.2cm, 0)
    node (tmp) {}
        -- (nmid -| tmp) -- (nmid.center);

    \draw [->] (lstm.north)  to node (smid){} (attn.south);
    \draw [->] (attn.north)  to node (nmid){} (resi2_pre.south);
    \draw [->] (resi2_pre.north)  -- (resi2.south);
    \draw[-] (smid.center) -- (smid -| lstm.west) -- ++ (-0.2cm, 0)
    node (tmp) {}
        -- (nmid -| tmp) -- (nmid.center);

    \node (a) at ($(conv.north) + (-\deltax, 0)$) {};
    \node (b) at ($(resi2.north) + (0, 0.5 * \yshift)$) {};
    \draw [-] (resi2.north) -- (b.center) -- (b-|a);


    \node (resi1) [resi] at ($(b.north) + (0, \yshift)$)
    {$\GELU(\LN(\Conv(C_{out}, C_{out} / 4, K=3, D=2)))$};

    \node (tmp) at ($ (b-|a) + (0, 0.5*\yshift) $) {};
    \draw [->] (tmp.center) -- (tmp -| resi1.south) -- (resi1.south);

    \node (lstm) [resi, yshift=\yshift, dashed] at (resi1.north)
    {$\LSTM(\mathrm{layers}=2, \mathrm{span}=200)$};

    \node (attn) [resi, yshift=\yshift, dashed] at (lstm.north)
    {$\mathrm{LocalAttention}(\mathrm{heads}=4)$};

    \node [anchor=north west] at (attn.north east) {$\Bigg\}$ if $i \in \{5, 6\}$};

    \node (resi2_pre) [resi, yshift=\yshift] at (attn.north)
    {$\GLU(\LN(\Conv(C_{out} / 4, 2 \cdot C_{out}, K=1)))$};
    \node (resi2) [resi, yshift=\yshift] at (resi2_pre.north)
    {$\mathrm{LayerScale}(\mathrm{init}=1\mathrm{e}{-}3)$};

    \draw [->] (resi1.north)  to node (smid){} (lstm.south);
    \draw[->] (lstm.north) to node (nmid){} (attn.south);
    \draw[-] (smid.center) -- (smid -| lstm.west) -- ++ (-0.2cm, 0)
    node (tmp) {}
        -- (nmid -| tmp) -- (nmid.center);

    \draw [->] (lstm.north)  to node (smid){} (attn.south);
    \draw [->] (attn.north)  to node (nmid){} (resi2_pre.south);
    \draw [->] (resi2_pre.north)  -- (resi2.south);
    \draw[-] (smid.center) -- (smid -| lstm.west) -- ++ (-0.2cm, 0)
    node (tmp) {}
        -- (nmid -| tmp) -- (nmid.center);

    \node (a) at ($(conv.north) + (-\deltax, 0)$) {};
    \node (b) at ($(resi2.north) + (0, 0.5 * \yshift)$) {};
    \draw [-] (resi2.north) -- (b.center) -- (b-|a);

    \node (rewrite) [rewrite, minimum width=\base - 2 * \dec, anchor=south, yshift=\yshift] at
      (conv.north |- resi2.north)  {$\GLU(\Conv(C_{out}, 2 \cdot C_{out}, K=1, S=1))$};
    \draw [->] ($(conv.north) + (-\deltax, 0)$) -- ($(rewrite.south) + (-\deltax, 0)$);

    \node (skip) [inout, anchor=south] at ($(rewrite.north) + (-\deltax, \yshift)$) {$\mathrm{Decoder}_i$};
    \draw[->] ($(rewrite.north) + (-\deltax, 0)$) -- (skip.south);

    \node (prev) [inout, anchor=north] at ($(conv.south) - (0, \yshift)$) {$\mathrm{Encoder}_{i-1}$ or input};
    \draw[->]  (prev.north) -- (conv.south);

    \node (next) [inout, anchor=south] at ($(rewrite.north) + (\deltax, \yshift)$) {$\mathrm{Encoder}_{i+1}$};
    \draw[->]  ($(rewrite.north) + (\deltax, 0)$) -- (next.south);
\end{tikzpicture}
}
\caption{Representation of the compressed residual branches that are
added to each encoder layer. For the 5th and 6th layer, a BiLSTM and a
local attention layer are added.}\label{fig:residual}
}
\end{figure}
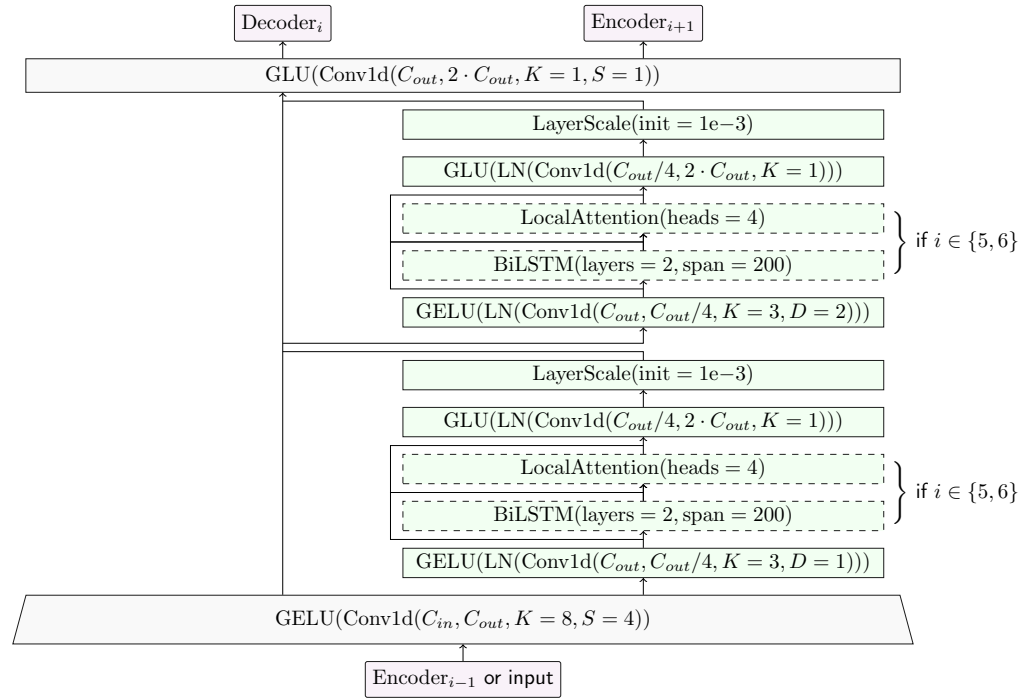

Finally, and for all layers, a final convolution with a kernel size of 1
outputs twice as many channels as the input dimension of the residual
branch, followed by a GLU. This output is then summed with the original
input, after having been scaled through a LayerScale layer
\citep{layerscale}, with an initial scale of \(1\mathrm{e}{-}3\). A
complete representation of the compressed residual branches is given on
\ref{fig:residual}.

\hypertarget{local-attention}{%
\subsubsection{Local attention}\label{local-attention}}

Local attention builds on regular attention \citep{attention} but
replaces positional embedding by a controllable penalty term that
penalizes attending to positions that are far away. Formally, the
attention weights from position \(i\) to position \(j\) is given by \[
w_{i, j} = \mathrm{softmax}(Q_i^T K_j - \sum_{k=1}^4 k \beta_{i, k} |i -j|) \]
where \(Q_i\) are the queries and \(K_j\) are the keys. The values
\(\beta_{i, k}\) are obtained as the output of a linear layer,
initialized so that they are initially very close to 0. Having multiple
\(\beta_{i, k}\) with different weights \(k\) allows the network to
efficiently reduce its receptive field without requiring
\(\beta_{i, k}\) to take large values. In practice, we use a sigmoid
activation to derive the values \(\beta_{i, k}\).

Interestingly, a similar idea has been developed in NLP~\citep{press2021train}, although
with a fixed penalty rather than a dynamic and learnt one done here.

\hypertarget{stabilizing-training}{%
\subsection{Stabilizing training}\label{stabilizing-training}}

We observed that Demucs training could be unstable, especially as we
added more layers and increased the training set size with 150 extra
songs. Loading the model just before its divergence point, we realized
that the weights for the innermost encoder and decoder layers would get
very large eigen values.

A first solution is to use group normalization (with 4 groups) just
after the non residual convolutions for the layers 5 and 6 of the
encoder and the decoder. Using normalization on all layers deteriorates
performance, but using it only on the innermost layers seems to
stabilize training without hurting performance. Interestingly, when the
training is stable (in particular when trained only on MusDB), using
normalization was at best neutral with respect to the separation score,
but never improved it, and considerably slowed down convergence during
the first half of the epochs. When the training was unstable, using
normalization would improve the overall performance as it allows the
model to train for a larger number of epochs.

A second solution we investigated was to use singular value
regularization \citep{spectral}. While previous work used the power
method iterative procedure, we obtained better and faster approximations
of the largest singular value using a low rank SVD method
\citep{lowranksvd}. This solution has the advantage of always improving
generalization, even when the training was already stable. Sadly, it was
not sufficient on its own to remove entirely instabilities, but only to
reduce them. Another down side was the longer training time due to the
extra low rank SVD evaluation. In the end, in order to both achieve the
best performance and remove entirely training instabilities, the two
solutions were combined.

\newpage

\hypertarget{experimental-results}{%
\section{Experimental Results}\label{experimental-results}}

\textbf{Important:} The results presented in this section are results obtained
as part of the MDX challenge. We provide easier to reproduce results and detailed ablation
that were conducted after the challenge in the Section "Reproducibility and Ablation" hereafter.

\hypertarget{datasets}{%
\subsection{Datasets}\label{datasets}}

The 2021 MDX challenge \citep{mdx} offered two tracks: Track A, where
only MusDB HQ \citep{musdbhq} could be used for training, and Track B,
where any data could be used. MusDB HQ, released under mixed
licensing\footnote{https://github.com/sigsep/website/blob/master/content/datasets/assets/tracklist.csv}
is composed of 150 tracks, including 86 for the train set, 14 for the
valid, and 50 for the test set. For Track B, we additionally trained
using 150 tracks for an internal dataset, and repurpose the test set of
MusDB as training data, keeping only the original validation set for
model selection. Models are evaluated either through the MDX AI Crowd
API\footnote{https://www.aicrowd.com/challenges/music-demixing-challenge-ismir-2021},
or on the MusDB HQ test set.

\hypertarget{realistic-remix-of-tracks}{%
\subsubsection{Realistic remix of
tracks}\label{realistic-remix-of-tracks}}

We achieved further gains, by fine tuning the
models on a specifically crafted dataset, and with longer training
samples (30 seconds instead of 10). This dataset was built by combining
stems from separate tracks, while respecting a number of conditions, in
particular beat matching and pitch compatibility. Note that training
from scratch on this dataset led to worse performance, likely because
the model could rely too much on melodic structure, while random
remixing forces the model to separate without this information.

We use \texttt{librosa} \citep{librosa} for both beat tracking and tempo
estimation, as well as chromagram estimation. Beat tracking is applied
only on the drum source, while chromagram estimation is applied on the
bass line. We aggregate the chromagram over time to a single chroma
distribution and find the optimal pitch shift between two stems to
maximize overlap (as measured by the L1 distance). We assume that the
optimal shift for the bass line is the same for the vocals and
accompaniments. Similarly, we align the tempo and first beat. In order
to limit artifacts, we only allow two stems to blend if they require
less than 3 semi-tones of shift and 15\% of tempo change.

\hypertarget{metrics}{%
\subsection{Metrics}\label{metrics}}

The MDX challenge introduced a novel Signal-To-Distortion measure.
Another SDR measure existed, as introduced by \citet{measures}. The
advantage of the new definition is its simplicty and fast evaluation.
The new definition is simply defined as \begin{equation}
SDR = 10 \log_{10} \frac{\sum_{n} \lVert\mathbf{s}(n)\rVert^2 + \epsilon}{\sum_{n}\lVert\mathbf{s}(n)-\hat{\mathbf{s}}(n)\rVert^2 + \epsilon},
\label{eq:sdr}
\end{equation} where \(s\) is the ground truth source, \(\hat{s}\) the
estimated source, and \(n\) the time index. In order to reliably compare
to previous work, we will refer to this new SDR definition as
\emph{nSDR}, and to the old definition as \emph{SDR}. Note that when
using nSDR on the MDX test set, the metric is defined as the average
across all songs. The evaluation on the MusDB test set follows the
traditional median across the songs of the median over all 1 second
segments of each song.

\hypertarget{models}{%
\subsection{Models}\label{models}}

The model submitted to the competitions were actually bags of 4 models.
For Track A, we had to mix hybrid and non hybrid Demucs models, as the
hybrid ones were having worse performance on the bass source. On Track
B, we used only hybrid models, as the extra training data allowed them
to perform better for all sources. Note that a mix of Hybrid models
using CaC or real masking were used, mostly because it was too
costly to reevaluate all models for the competition. For details on the
exact architecture and hyper-parameter used, we refer the reader to our
Github repository
\href{https://github.com/facebookresearch/demucs}{facebookresearch/demucs}.

For the baselines, we report the numbers from the top participants at
the MDX competition \citep{mdx}. We focus particularly on the
KUIELAB-MDX-Net model, which came in second. This model builds on
\citep{cac} and combines a pure spectrogram model with the prediction
from the original Demucs \citep{demucs} model for the drums and bass
sources. When comparing models on MusDB, we also report the numbers for
some of the best performing methods outside of the MDX competition,
namely D3Net \citep{d3net} and ResUNetDecouple+
\citep{kong2021decoupling}, as well as the original Demucs model
\citep{demucs}. Note that those models were evaluated on MusDB (not HQ)
which lacks the frequency content between 16 kHz and 22kHz. This can
bias the metrics.

\hypertarget{results-on-mdx}{%
\subsection{Results on MDX}\label{results-on-mdx}}

We provide the results from the top participants at the MDX competition
on Table \ref{tbl:mdx_a} for the track A (trained on MusDB HQ only)
and on Table \ref{tbl:mdx_b} for track B (any training data). We
also report for track A the metrics for the Demucs architecture improved
with the residual branches, but without the spectrogram branch. The
hybrid approach especially improves the nSDR on the \texttt{Other} and
\texttt{Vocals} source. Despite this improvement, the Hybrid Demucs
model is still performing worse than the KUIELAB-MDX-Net on those two
sources. On Track B, we notice again that the Hybrid Demucs architecture
is very strong on the \texttt{Drums} and \texttt{Bass} source, while
lagging behind on the \texttt{Other} and \texttt{Vocals} source.

\begin{longtable}[]{@{}llllll@{}}
\caption{Results of Hybrid Demucs on the MDX test set, when trained only
on MusDB (track A) using the nSDR metric. \label{tbl:mdx_a}}\tabularnewline
\toprule
Method & \texttt{All} & \texttt{Drums} & \texttt{Bass} & \texttt{Other}
& \texttt{Vocals}\tabularnewline
\midrule
\endfirsthead
\toprule
Method & \texttt{All} & \texttt{Drums} & \texttt{Bass} & \texttt{Other}
& \texttt{Vocals}\tabularnewline
\midrule
\endhead
Hybrid Demucs & \textbf{7.33} & \textbf{8.04} & \textbf{8.12} & 5.19 &
7.97\tabularnewline
KUIELAB-MDX-Net & 7.24 & 7.17 & 7.23 & \textbf{5.63} &
\textbf{8.90}\tabularnewline
Music\_AI & 6.88 & 7.37 & 7.27 & 5.09 & 7.79\tabularnewline
\bottomrule
\end{longtable}

\begin{longtable}[]{@{}llllll@{}}
\caption{Results of Hybrid Demucs on the MDX test set, when trained with
extra training (track B) using the nSDR metric.
\label{tbl:mdx_b}}\tabularnewline
\toprule
Method & \texttt{All} & \texttt{Drums} & \texttt{Bass} & \texttt{Other}
& \texttt{Vocals}\tabularnewline
\midrule
\endfirsthead
\toprule
Method & \texttt{All} & \texttt{Drums} & \texttt{Bass} & \texttt{Other}
& \texttt{Vocals}\tabularnewline
\midrule
\endhead
Hybrid Demucs & 8.11 & \textbf{8.85} & \textbf{8.86} & 5.98 &
8.76\tabularnewline
KUIELAB-MDX-Net & 7.37 & 7.55 & 7.50 & 5.53 & 8.89\tabularnewline
AudioShake & \textbf{8.33} & 8.66 & 8.34 & \textbf{6.51} &
\textbf{9.79}\tabularnewline
\bottomrule
\end{longtable}

\hypertarget{results-on-musdb}{%
\subsection{Results on MusDB}\label{results-on-musdb}}

We show on Table \ref{tbl:musdb} the SDR metrics as measured on the
MusDB dataset. Again, Hybrid Demucs achieves the best performance for
the \texttt{Drums} and \texttt{Bass} source, while improving quite a lot
over waveform only Demucs for the \texttt{Other} and \texttt{Vocals},
but not enough to surpasse KUIELAB-MDX-Net, which is purely spectrogram
based for those two sources. Interestingly, the best performance on the
\texttt{Vocals} source is also achieved by ResUNetDecouple+
\citep{kong2021decoupling}, which uses a novel complex modulation of the
input spectrogram.

\hypertarget{human-evaluations}{%
\subsection{Human evaluations}\label{human-evaluations}}

We also performed Mean Opinion Score human evaluations. We re-use the
same protocol as in \citep{demucs}: we asked human subjects to evaluate
a number of samples based on two criteria: the absence of artifacts, and
the absence of bleeding (contamination). Both are evaluated on a scale
from 1 to 5, with 5 being the best grade. Each subject is tasked with
evaluating 25 samples of 12 seconds, drawn randomly from the 50 test set
tracks of MusDB. All subjects have a strong experience with music
(amateur and professional musicians, sound engineers etc). The results
are given on Table \ref{tbl:mos_quality} for the quality, and
\ref{tbl:mos_bleed} for the bleeding. We observe strong improvements
over the original Demucs, although we observe some regression on the
bass source when considering quality. The model KUIELAb-MDX-Net that
came in second at the MDX competition performs the best on vocals. The
Hybrid Demucs architecture however reduces by a large amount bleeding
across all sources.

\begin{longtable}[]{@{}lllllll@{}}
\caption{Comparison on the MusDB (HQ for Hybrid Demucs) test set, using
the original SDR metric. This includes methods that did not participate
in the competition. ``Mode'' indicates if the waveform (W) or spectrogram
(S) domain is used. Model with a ``*'' were evaluated on MusDB HQ.
\label{tbl:musdb}}\tabularnewline
\toprule
Method & Mode & \texttt{All} & \texttt{Drums} & \texttt{Bass} &
\texttt{Other} & \texttt{Vocals}\tabularnewline
\midrule
\endfirsthead
\toprule
Method & Mode & \texttt{All} & \texttt{Drums} & \texttt{Bass} &
\texttt{Other} & \texttt{Vocals}\tabularnewline
\midrule
\endhead
Hybrid Demucs* & S+W & \textbf{7.68} & \textbf{8.24} & \textbf{8.76} &
5.59 & 8.13\tabularnewline
Demucs v2 & W & 6.28 & 6.86 & 7.01 & 4.42 & 6.84\tabularnewline
KUIELAB-MDX-Net* & S+W & 7.47 & 7.20 & 7.83 & \textbf{5.90} &
\textbf{8.97}\tabularnewline
D3Net & S & 6.01 & 7.01 & 5.25 & 4.53 & 7.24\tabularnewline
ResUNetDecouple+ & S & 6.73 & 6.62 & 6.04 & 5.29 &
\textbf{8.98}\tabularnewline
\bottomrule
\end{longtable}

\begin{longtable}[]{@{}llllll@{}}
\caption{Mean Opinion Score results when asking to rate the quality and
absence of artifacts in the generated samples, from 1 to 5 (5 being the
best grade). Standard deviation is around 0.15.
\label{tbl:mos_quality}}\tabularnewline
\toprule
Method & \texttt{All} & \texttt{Drums} & \texttt{Bass} & \texttt{Other}
& \texttt{Vocals}\tabularnewline
\midrule
\endfirsthead
\toprule
Method & \texttt{All} & \texttt{Drums} & \texttt{Bass} & \texttt{Other}
& \texttt{Vocals}\tabularnewline
\midrule
\endhead
Ground Truth & 4.12 & 4.12 & 4.25 & 3.92 & 4.18\tabularnewline
Hybrid Demucs & \textbf{2.83} & \textbf{3.18} & 2.58 & \textbf{2.98} &
2.55\tabularnewline
KUIELAB-MDX-Net & \textbf{2.86} & 2.70 & 2.68 & \textbf{2.99} &
\textbf{3.05}\tabularnewline
Demucs v2& 2.36 & 2.62 & \textbf{2.89} & 2.31 &
1.78\tabularnewline
\bottomrule
\end{longtable}

\begin{longtable}[]{@{}llllll@{}}
\caption{Mean Opinion Score results when asking to rate the absence of
bleeding between the sources, from 1 to 5 (5 being the best grade).
Standard deviation is around 0.15. \label{tbl:mos_bleed}}\tabularnewline
\toprule
Method & \texttt{All} & \texttt{Drums} & \texttt{Bass} & \texttt{Other}
& \texttt{Vocals}\tabularnewline
\midrule
\endfirsthead
\toprule
Method & \texttt{All} & \texttt{Drums} & \texttt{Bass} & \texttt{Other}
& \texttt{Vocals}\tabularnewline
\midrule
\endhead
Ground Truth & 4.40 & 4.51 & 4.52 & 4.13 & 4.43\tabularnewline
Hybrid Demucs & \textbf{3.04} & \textbf{2.95} & \textbf{3.25} &
\textbf{3.08} & \textbf{2.88}\tabularnewline
KUIELAB-MDX-Net & 2.44 & 2.23 & 2.19 & 2.64 & 2.66\tabularnewline
Demucs v2 & 2.37 & 2.24 & 2.96 & 1.99 & 2.46\tabularnewline
\bottomrule
\end{longtable}

\hypertarget{repro}{%
\subsection{Reproducibility and Ablation}\label{repro}}

In this section, we provide ablation of the performance of the model,
as well as a simpler setup for reproducing the performance of the model submitted to MDX Track A.
Note that the numbers and analysis presented here might differ slightly from the ones presented up to now,
and should be preferred when referring to this work.

\subsubsection{Reproducibility}

The model submitted to the MDX competition Track A
used heterogeneous configurations, as we used any model that was sufficiently trained
at any given time. This led to a complex bag of 4 models,
some of which were used only for some sources. While suitable for a competition, such a complex
model does get in the way of reproducing easily the performance achieved.

The training grids for all the models presented in this section can be found
on GitHub repository \href{https://github.com/facebookresearch/demucs}{facebookresearch/demucs},
in the file \texttt{demucs/grids/repro.py}, and \texttt{demucs/grids/repro\_ft.py} for the fine tuned
models.

We reproduced the performance of the model submitted to Track A by training two time domain Demucs
(with the residual branches depicted on Figure~\ref{fig:residual}), and two hybrid Demucs
using Complex-As-Channels representation. All four models were trained for 600 epochs,
using the SVD penalty and exponential moving average on the weights. Within each domain, only
the random seed changes between the two models.
The four models were fined tuned on the realistic remix of tracks dataset.
The predictions of the four models are averaged into the final prediction, with equal weights
over all sources.

As a first ablation, we also form one bag composed of the two time domain only models, and another
bag with the two hybrid models only. For fairness, we evaluate each model twice with a random shift,
which is known to improve the performance~\citep{demucs}. We also compare to the original Demucs
model, retrained on MusDB HQ, using 10 random shifts, as done in~\citep{demucs}.

The results are presented on Table~\ref{tbl:repro}. We reach an overall SDR of 7.64 dB, just 0.04 dB
of the model submitted to MDX Track A. We notice a difference in performance between time only,
and hybrid only bags only for the \texttt{bass} source and the \texttt{other} source. We
still decided to use the same weights over all sources for each type of model for simplicity.
We can see the advantage of averaging multiple models, as the combination of both time only
and hybrid only model surpasses either ones individually for instance on the \texttt{drums} or the \texttt{vocals} sources.
Note however that when training with extra training data, e.g. for the MDX Track B models,
the hybrid models were always better than the time only ones.

\begin{longtable}[]{@{}lllllll@{}}
\caption{Comparison on the MusDB HQ test set, using
the original SDR metric of different bags of models, as well as with
the original Demucs v2 model retrained on MusDB HQ. ``Mode'' indicates if the waveform (W) or spectrogram
(S) domain is used.
\label{tbl:repro}}\tabularnewline
\toprule
Method & Mode & \texttt{All} & \texttt{Drums} & \texttt{Bass} &
\texttt{Other} & \texttt{Vocals}\tabularnewline
\midrule
\endfirsthead
\toprule
Method & Mode & \texttt{All} & \texttt{Drums} & \texttt{Bass} &
\texttt{Other} & \texttt{Vocals}\tabularnewline
\midrule
\endhead
Bag time + hybrid& S+W & \textbf{7.64} & \textbf{8.12} & \textbf{8.43} &
\textbf{5.65} & \textbf{8.35}\tabularnewline
Bag time only & W & 7.27 & 7.57 & \textbf{8.38} & 5.17 & 7.96\tabularnewline
Bag hybrid only & S+W & 7.34 & 7.96 & 7.85 & \textbf{5.63} & 7.95\tabularnewline
Demucs v2 HQ & W & 6.17 & 6.54 & 7.08 & 4.21 & 6.85\tabularnewline
\bottomrule
\end{longtable}

\subsubsection{Ablation}

We report on Table~\ref{tbl:ablation} a short ablation study of the model.
We start from a time only improved Demucs, e.g. trained with residual
branches, local attention and svd penalty. We can first oberve the effect
of fine tuning on a set of realistic remixes, which improves by 0.3 dB the SDR overall.
Further gains are achieved using the bagging. Using Exponential Moving Average on the weights
improves the SDR by 0.2dB. The effect of the SVD penalty is more contrasted, with on overall gain of 0.1dB,
mainly due to the improved \texttt{vocals} (+0.7 dB), but with a deterioration
on the drums source (-0.4 dB).

Finally, removing the LSTM or the local attention in the residual branches lead to a strong
decrease of the SDR. Interestingly, the local attention is the most important, despite the absence
of positional embedding.
One decision taken during the challenge was to switch to GELU instead of ReLU. The ablation indicates that no real gain is achieved here.

\begin{longtable}[]{@{}llllll@{}}
\caption{Ablation study, all models are trained and evaluated on MusDB HQ.
The base model is a time only improved Demucs,
with local attention, residual branches and svd penalty.
Note that we report single model performance instead of bags of model.
\label{tbl:ablation}}\tabularnewline
\toprule
Model & \texttt{All} & \texttt{Drums} & \texttt{Bass} &
\texttt{Other} & \texttt{Vocals}\tabularnewline
\midrule
\endfirsthead
\toprule
Model & \texttt{All} & \texttt{Drums} & \texttt{Bass} &
\texttt{Other} & \texttt{Vocals}\tabularnewline
\midrule
\endhead
Improved Time Demucs & 6.83 & 7.06 & 7.78 & 4.81 & 7.65\tabularnewline
\textbf{+} fine tuning & 7.11 & 7.42 & 8.18 & 5.08 & 7.75\tabularnewline
\textbf{+} fine tuning and bagging & 7.27 & 7.57 & 8.38 & 5.17 & 7.96\tabularnewline
\textbf{-} LSTM in branch & 6.44 & 6.66 & 6.68 & 4.89 & 7.54\tabularnewline
\textbf{-} Local Attention & 6.29 & 6.39 & 6.76 & 4.68 & 7.33\tabularnewline
\textbf{-} SVD penalty & 6.73 & 7.45 & 8.01 & 4.48 & 6.98\tabularnewline
\textbf{-} EMA on weights & 6.63 & 6.99 & 7.36 & 4.74 & 7.43\tabularnewline
\textbf{-} GELU \textbf{+} ReLU & 6.84 & 7.19 & 7.81 & 4.73 & 7.63\tabularnewline
\bottomrule
\end{longtable}

\hypertarget{conclusion}{%
\section{Conclusion}\label{conclusion}}

We introduced a number of architectural changes to the Demucs
architecture that greatly improved the quality of source separation for
music. On the MusDB HQ benchark, the gain is around 1.4 dB. Those
changes include compressed residual branches with local attention and
chunked biLSTM, and most importantly, a novel hybrid
spectrogram/temporal domain U-Net structure, with parallel temporal and
spectrogram branches, that merge into a common core. Those changes
allowed to achieve the first rank at the 2021 Sony Music DemiXing
challenge, and translated into strong improvements of the overall
quality and absence of bleeding between sources as measured by human
evaluations. For all its gain, one limitation of our approach is the
increased complexity of the U-Net encoder/decoder, requiring careful
alignmement of the temporal and spectral signals through well shaped
convolutions.

\clearpage

\bibliography{paper}

\end{document}